\documentstyle [12pt,epsf] {article}

\catcode`@=12
\newcommand{\beq}{\begin{equation}}
\newcommand{\eeq}{\end{equation}}

\newcommand{\bea}{\begin{eqnarray}}
\newcommand{\eea}{\end{eqnarray}}

\begin{document}
\thispagestyle{empty}
\begin{flushright} UCRHEP-T283\\July 2000\
\end{flushright}
\vspace{0.5in}
\begin{center}
{\Large \bf Low-Scale Axion from Large Extra Dimensions\\}
\vspace{1.5in}
{\bf Ernest Ma$^1$, Martti Raidal$^{1,2}$, and Utpal Sarkar$^{1,3}$\\}
\vspace{0.2in}
{$^1$ \sl Department of Physics, University of California, Riverside, 
California 92521, USA\\}
\vspace{0.1in}
{$^2$ \sl National Insitute of Chemical and Biological Physics, Tallinn 10143, 
Estonia\\}
\vspace{0.1in}
{$^3$ \sl Physical Research Laboratory, Ahmedabad 380 009, India\\}
\vspace{1.5in}
\end{center}
\begin{abstract}

The mass of the axion and its decay rate are known to depend only on the 
scale of Peccei-Quinn symmetry breaking, which is constrained by astrophysics 
and cosmology to be between $10^9$ and $10^{12}$ GeV.  We propose a new 
mechanism such that this effective scale is preserved and yet the fundamental 
breaking scale of $U(1)_{PQ}$ is very small (a kind of inverse seesaw) in 
the context of large extra dimensions with an anomalous U(1) gauge symmetry 
in our brane. Unlike any other (invisible) axion model, there are now 
possible collider signatures in this scenario.
\end{abstract}

\newpage
\baselineskip 24pt

Although CP violation has been observed in weak interactions \cite{cp1,cp2} 
and it is required for an explanation of the baryon asymmetry of the 
universe \cite{asym}, it becomes a problem in strong interactions.  The 
reason is that the multiple vacua of quantum chromodynamics (QCD) connected by 
instantons \cite{insta} require the existence of the CP violating $\theta$ 
term \cite{theta}
\begin{equation}
{\cal L}_\theta = \theta_{QCD} {g_s^2 \over 32 \pi^2} 
G_{\mu \nu}^a \widetilde G^{a \mu \nu} ,
\end{equation}
where $g_s$ is the strong coupling constant, $G^a_{\mu \nu}$ is the gluonic 
field strength and $\tilde G^a_{\mu \nu}$ is its dual. Nonobservation of 
the electric dipole moment of the neutron \cite{edm} implies that
\begin{equation}
\bar \theta = \theta_{QCD} - Arg ~Det ~M_u ~M_d < 10^{-10},
\end{equation}
instead of the theoretically expected order of unity.  In the above, $M_u$ 
and $M_d$ are the respective mass matrices of the charge 2/3 and $-1/3$ 
quarks of the standard model of particle interactions.  This is commonly 
known as the strong CP problem. 

The first and best motivated solution to the strong CP problem was proposed 
by Peccei and Quinn \cite{pq}, in which the quarks acquire a dynamical phase 
from the spontaneous breaking of a new global symmetry [$U(1)_{PQ}$] and 
relaxes $\bar \theta$ to its natural minimun value of zero.  As a result, 
there appears a Goldstone boson called the axion but it is not strictly 
massless \cite{ww} because it couples to two gluons (like the neutral pion) 
through the axial triangle anomaly \cite{anomal}.

The scale of $U(1)_{PQ}$ breaking (which is conventionally identified with 
the axion decay constant $f_a$) determines the axion coupling to gluons, 
which is proportional to $1/f_a$.  If $f_a$ is the electroweak symmetry 
breaking scale as originally proposed \cite{pq}, then the model is already 
ruled out by laboratory experiments \cite{expt}.  
In fact, $f_a$ is now known to be constrained by astrophysical and 
cosmological arguments \cite{astro} to be between $10^9$ and $10^{12}$ GeV. 
Hence the axion must be an electroweak singlet or predominantly so.  
It may couple to the usual quarks and leptons through a suppressed mixing 
with the standard Higgs doublet \cite{dfsz}, or it may couple only to other 
unknown colored fermions \cite{ksvz}, or it may couple to gluinos 
\cite{dms} as well as all other supersymmetric particles.

Because the axion must necessarily mix with the $\pi$ and $\eta$ mesons, 
it must have a two-photon decay mode.  This is the basis of all experimental 
attempts \cite{expt} to discover its existence.  On the other hand, the 
accompanying new particles in all viable axion models to date are very heavy, 
i.e. of order $f_a$; hence they are completely inaccessible to experimental 
verification.

In the following we consider instead the possiblily that the $U(1)_{PQ}$ 
breaking scale is actually very small, but that $f_a$ is large because of a 
kind of inverse seesaw mechanism.  We show how this scenario may be realized 
in the context of large extra dimensions with an anomalous U(1) gauge symmetry 
in our brane.  The associated new physics now exists at around 1 TeV, with 
a number of interesting observable consequences at future colliders.

We assume a singlet scalar field $\chi$ with a nonzero PQ charge existing in 
the bulk of large extra dimensions \cite{extra}.  The $shining$ \cite{distant} 
of this field in our brane is the source of spontaneous $U(1)_{PQ}$ breaking 
in our world (called a 3-brane).  The idea is that $\chi$ gets a large vacuum 
expectation value (VEV) in a distant brane, but its effect on our brane is 
small because we are far away from it.  (In the case of lepton number, this 
mechanism has been used recently to obtain small Majorana neutrino masses 
\cite{extnu}.)  To convert this small $\langle \chi \rangle$ to a large 
$f_a$, we need to assume an anomalous U(1) gauge symmetry in our brane at 
the TeV energy scale, as explained below.

In a theory of large extra dimensions with quantum gravity at the TeV scale, 
there is no large scale available for the axion.  Since the behavior of 
Goldstone bosons depends not on the coupling but only on the scale of 
symmetry breaking in general, it is a problem which is not easily resolved 
\cite{others}.  Here we find a new and novel solution to this apparent 
contradiction in the case where there is an anomalous U(1) gauge symmetry, 
which is of course well studied \cite{u1} as a possible manifestation of 
string theory near the string scale (now considered also at around a few TeV) 
and has well-known applications in quark and lepton Yukawa textures and 
supersymmetry breaking.

We extend the standard model of particle interactions to include an extra 
$U(1)_A$ gauge symmetry and an extra $U(1)_{PQ}$ global symmetry.  All 
standard-model particles are trivial under these two new symmetries.  We 
then introduce a new heavy quark singlet $\psi$ and two scalar singlets 
$\sigma$ and $\eta$ with $U(1)_A$ and $U(1)_{PQ}$ charges as shown in 
Table 1.  All fields except $\chi$ are confined to our brane.

\begin{table}[htb]
\caption{Peccei-Quinn charges of the fermions and scalars}
\begin{center}
\begin{tabular}{||c|c|c|c||}
\hline \hline
&\multicolumn{3}{c||}{ Transformation under} \\
\cline{2-4} 
Fields &$SU(3)_C \times SU(2)_L \times U(1)_Y$& $U(1)_A$&$U(1)_{PQ}$ \\
\hline
&&&\\
$(u_i, d_i)_L$ & (3,2,1/6) & 0 & 0 \\
$u_{iR}$ & (3,1,2/3) & 0 & 0  \\
$d_{iR}$ & (3,1,$-$1/3) & 0 & 0  \\
$(\nu_i, e_i)_L$ & (1,2,$-$1/2) & 0& 0  \\
$e_{iR} $& (1,1,$-$1) & 0 & 0  \\
&&&\\
\hline
&&&\\
$\psi_L$& (3,1,--1/3) & 1 & $k$ \\
$\psi_R$& (3,1,--1/3) & --1 & $-k$ \\
&&&\\
\hline
&&&\\
$(\phi^+, \phi^0)$ & (1,2,1/2)& 0 & 0  \\
$\sigma$ & (1,1,0) & 2 & $2k$ \\
$\eta $ & (1,1,0)& 2 & $2k-2$ \\
$\chi$ & (1,1,0) & 0 & 2 \\
&&&\\
\hline \hline
\end{tabular}
\end{center}
\end{table}

Because of our chosen charge assignments, only the field $\sigma$ couples 
to the colored fermion $\psi$, i.e.
\begin{equation}
{\cal L}_Y = f \sigma \bar \psi_L \psi_R + h.c.
\end{equation}
Hence it also couples to two gluons through 
the usual triangular loop.  As $\sigma$ acquires a VEV, say $u$, of order 
1 TeV, both $U(1)_A$ and $U(1)_{PQ}$ are broken, whereas the latter is broken 
by $\langle \chi \rangle = z$, and it induces a VEV also for $\eta$, i.e. 
$\langle \eta \rangle = w$.  We will show in the following that given $z$ is 
small from its origin in the bulk, $w$ is also small.  Now the longitudinal 
component of the $Z_A$ boson is mostly given by Im$\sigma$, so the axion is 
excluded to be mostly a linear combination of Im$\eta$ and Im$\chi$, 
but the latter two fields do not couple to the colored fermion $\psi$. 
As a result, the axion's coupling to two gluons is now effectively
\begin{equation}
{1 \over f_a} = {w^2 \over u^2 \sqrt {w^2 + z^2}},
\end{equation}
which can be thought of as a kind of inverse seesaw, i.e. the largeness 
of $f_a$ is explained by the smallness of $w$.  Details will be given later.

Our brane ${\cal P}$ is located at a point $y=0$ in the bulk.  Peccei-Quinn 
symmetry is broken maximally in a distant brane ${\cal P}'$, located at a 
point $y=y_*$ in the bulk.  We assume for simplicity that the separation of 
the two branes is of order the radius of compactification of the extra 
space dimensions, i.e. $|y_*|=r$, which is only a few $\mu$m in magnitude. 
The fundamental scale $M_*$ in this theory is then related to the reduced 
Planck scale $M_P = 2.4 \times 10^{18}$ GeV by the relation
\begin{equation}
r^n M_*^{n + 2} \sim M_P^2  .
\end{equation}
The $U(1)_{PQ}$ symmetry breaking in the distant brane acts as a point 
source $J$, which induces an effective VEV, i.e. $z$, to the singlet bulk 
field $\chi$.  Other effects which may perturb the $shined$ value of $\langle 
\chi \rangle$ in our world are all included as boundary conditions to the 
source $J$, so that the effect of the field $\chi$ in our brane always 
appears in the combination $z(y=0) e^{i\varphi}$, where $\varphi(x)$ is a 
dynamical phase which transforms under $U(1)_{PQ}$ to preserve its 
invariance.  This formulation has also been used for the spontaneous 
breaking of lepton number in the case of neutrinos \cite{extnu}.

In our brane, the profile of $\chi$ is given by the Yukawa potential
in the transverse dimensions 
\begin{equation}
\langle \chi(y = 0) \rangle = J(y=y_*) \Delta_n(r),
\end{equation}
where 
\begin{equation}
\Delta_n(r) = {1 \over (2 \pi )^{n \over 2}
M_*^{n- 3}} ~\left( {m_\chi \over r} \right)^{n-2 \over 2}
~K_{n - 2 \over 2} \left( m_\chi r \right),
\end{equation}
$K$ being the modified Bessel function.  We consider the source
to be dimensionless, which we take to be $J=1$.  For the interesting case 
of $n> 2$ and $m_\chi r \ll 1$, the $shined$ value of $\chi$ is given by
\begin{equation}
\langle \chi \rangle \approx \displaystyle{
{ \Gamma ( {n -2 \over 2} ) \over
4 \pi^{n \over 2} }{M_* \over (M_* r)^{n-2} } }  
= \displaystyle{
{ \Gamma ( {n -2 \over 2} ) \over
4 \pi^{n \over 2} }~ M_* ~\left({M_* \over M_P}\right)^{2 - (4/n)} }.
\end{equation}
For $n=3$ and $M_* = 10$ TeV, we get $\langle \chi \rangle \sim 0.2$ keV.  
This is the smallest value possible with our assumptions.  However, if 
the distant brane is located at $y_*$ less than $r$, larger values of 
$\langle \chi \rangle$ may be obtained.  As we will show, the range 1 keV 
to 1 MeV corresponds nicely to the axion decay constant of $10^{12}$ to 
$10^9$ GeV.

We express the bulk field as
\begin{equation}
\chi = {1 \over \sqrt 2} ( \rho + z \sqrt 2) e^{i \varphi} .
\end{equation}
Its self-interaction terms are now given by
\begin{equation}
V(\chi) = \lambda z(y)^2 \rho(x,y)^2 + {1 \over \sqrt 2} \lambda
z(y) \rho(x,y)^3 + {1 \over 8} \lambda \rho(x,y)^4 .
\end{equation}
This Lagrangian has the virtue of universality, i.e., $\lambda$ is unchanged,
but $z$ can change depending on where our brane is from the distant
brane.  The invariance under $U(1)_{PQ}$, i.e. $\rho \to \rho$ and $\varphi 
\to \varphi + 2 \theta$, is also maintained in the other interactions, as 
described below.  The parameters in the potential of $\chi$ are thus 
guaranteed to be independent of the parameters of our brane. 

The scalar potential in our brane excluding $V(\chi)$ is now given by
\begin{eqnarray}
V &=& m_1^2 \Phi^\dagger \Phi + m_2^2 \sigma^\dagger \sigma + m_3^2 
\eta^\dagger \eta + {1 \over 2} \lambda_1 (\Phi^\dagger \Phi)^2 + {1 \over 2} 
\lambda_2 (\sigma^\dagger \sigma)^2 + {1 \over 2} \lambda_3 (\eta^\dagger 
\eta)^2 \nonumber \\ && + \lambda_4 (\Phi^\dagger \Phi)(\sigma^\dagger \sigma) 
+ \lambda_5 (\Phi^\dagger \Phi)(\eta^\dagger \eta) + \lambda_6 
(\sigma^\dagger \sigma)(\eta^\dagger \eta) + (\mu z e^{i \varphi} 
\sigma^\dagger \eta + h.c.),
\end{eqnarray}
where $\mu$ has the dimension of mass and we assume that all mass 
parameters are of the same order of magnitude, i.e. 1 TeV.

The minimum of $V$ satisfies the following conditions:
\begin{eqnarray}
m_1^2 + \lambda_1 v^2 + \lambda_4 u^2 + \lambda_5 w^2 &=& 0, \\
u(m_2^2 + \lambda_2 u^2 + \lambda_4 v^2 + \lambda_6 w^2) + \mu z w &=& 0, \\
w(m_3 ^2 + \lambda_3 w^2 + \lambda_5 v^2 + \lambda_6 u^2) + \mu z u  &=& 0,
\end{eqnarray}
where $\langle \phi^0 \rangle = v$. Hence
\begin{eqnarray}
v^2 &\simeq& {-\lambda_2 m_1^2 + \lambda_4 m_2^2 \over \lambda_1 \lambda_2 - 
\lambda_4^2}, \\ u^2 &\simeq& {-\lambda_1 m_2^2 + \lambda_4 m_1^2 \over 
\lambda_1 \lambda_2 - \lambda_4^2},
\end{eqnarray}
and
\begin{equation}
w \simeq {- \mu z u \over m_3^2 + \lambda_5 v^2 + \lambda_6 u^2},
\end{equation}
which is indeed of order $z$ as mentioned earlier.

Whereas Im$\phi^0$ becomes the longitudinal component of the usual $Z$ boson, 
$(u {\rm Im} \sigma + w {\rm Im} \eta)/\sqrt {u^2 + w^2}$ becomes that of the 
new $Z_A$ boson.  Since the $3 \times 3$ mass matrix in the basis [Im$\sigma$, 
Im$\eta$, $z\varphi$] is given by
\begin{equation}
\pmatrix{-\mu z w / u 
& \mu z& \mu w \cr 
\mu z & - \mu z u / w &
-\mu u \cr \mu w &  -\mu u & - \mu u w / z },
\end{equation}
the axion $a$ is identified as the following:
\begin{eqnarray} 
{a \over \sqrt 2} &=& {1 \over {N}} \left[ uw^2 {\rm Im} \sigma  
- w u^2 {\rm Im} \eta + z (u^2 + w^2) z \varphi \right] \nonumber \\
&\simeq& {w^2 \over u \left( w^2 + z^2 \right)^{1/2} } {\rm Im} \sigma -
{w \over \left( w^2 + z^2 \right)^{1/2} } {\rm Im} \eta + 
{z \over \left( w^2 + z^2 \right)^{1/2} } z \varphi,
\end{eqnarray}
where $N= \left\{ w^2 u^2 (w^2 + u^2) + z^2 (w^2 + u^2)^2 \right\}^{1/2}$ is 
the normalization.  Since only $\sigma$ couples to the colored fermion $\psi$ 
and the component of Im$\sigma$ in the axion is $u$ times a phase, the axion 
coupling to the gluons through $\psi$ is effectively as given by Eq.~(4) as 
mentioned earlier.  Using $u \sim 1$ TeV and $w \sim z \sim 1$ keV to 1 MeV, 
we see that $f_a$ is indeed in the range $10^{12}$ to $10^9$ GeV.

In Table 1, we have not specified the value of $k$ for the PQ charge of 
$\psi$.  This is intentional because our model is independent of it.  This 
ambiguity also helps us to understand its pattern of symmetry breaking. 
For example, if $\langle \chi \rangle = 0$, then $\langle \eta \rangle = 0$ 
also.  In that case, there is no axion and the Peccei-Quinn symmetry 
disappears, i.e. $k=0$.  Hence the true scale of $U(1)_{PQ}$ breaking is 
indeed small, i.e. $z$ from the bulk, as asserted.

To understand why we have an exception to the general rule that the axion 
coupling is inversely proportional to the scale of $U(1)_{PQ}$ breaking, we 
point out that the anomalous nature of $U(1)_A$ is crucial.  If we attempt 
to make it free of the axial triangle anomaly, we need to add colored 
fermions with opposite $U(1)_A$ charges to $\psi_{L,R}$.  They must then 
acquire mass through a new scalar field with opposite $U(1)_A$ charge to 
$\sigma$.  The longitudinal component of $Z_A$ takes up a linear combination 
of the two imaginary parts, leaving free the other to be the axion, which now 
couples to the colored fermions with the same scale as $U(1)_A$ symmetry 
breaking. The above is of course the analog of what happens in the well-known 
original Peccei-Quinn proposal \cite{pq}.

All axion models to date have no accompanying verifiable new physics other 
than the $a \to \gamma \gamma$ decay, and that depends on the axion being 
a component of dark matter.  In our scenario, the possibility exists for 
this new physics to be at the TeV scale and be observable at future colliders.

(1) The stable heavy colored fermion $\psi$ may be produced in pairs, i.e. 
$gg \to \psi \bar \psi$.  Both $\psi$ and $\bar \psi$ carry light quarks and 
gluons with them and appear as jets, but when these jets hit the hadron 
calorimeter in a typical detector, a large part (i.e. 2$m_\psi$) of the 
initial collision energy is ``frozen'' in the mass and appears ``lost''.

(2) There is mixing between the standard-model Higgs boson Re$\phi^0$ with 
the new scalar Re$\sigma$ of order $v/u$, i.e. 0.1 or so.  This means 
that the lighter (call it $h$) of the two physical scalar bosons has a 
small component of Re$\sigma$, but that only modifies its (small) $gg$ 
and $\gamma \gamma$ decay amplitudes through the $\psi$ loop.  Hence $h$ 
behaves almost exactly like the standard-model Higgs boson.

(3) The $U(1)_A$ gauge boson $Z_A$ may be produced by $gg^*$ fusion through 
the $\psi$ loop.  If kinematically allowed, it will decay into Re$\eta$ + 
Im$\eta$.  Since Im$\eta$ is partly ($w/\sqrt{w^2+z^2}$) the axion $a$ 
which will escape detection, this event has a lot of possible missing 
transverse momentum.  The subsequent decay of Re$\eta$ is into $a$ and a 
virtual $Z_A$ which turns into $gg$.  This adds more missing transverse 
momentum.  The end result of the production and subsequent decay of $Z_A$ 
is thus two gluon jets and two axions.  This is a distinctive signature of 
our scenario \cite{note}.  It predicts collider events with large missing 
energy without the existence of supersymmetry.

{\it Acknowledgement.} This work was supported in part by the U.~S.~Department
of Energy under Grant No.~DE-FG03-94ER40837.  One of us (U.S.) would 
like to thank the Physics Department, University of California, 
Riverside for hospitality.

\newpage
\bibliographystyle{unsrt}

\begin{thebibliography}{99}
\bibitem{note} Re$\eta$ will also decay (through its mixing with Re$\phi^0$) 
into standard-model final states such as $ZZ$, $WW$, etc.  However, this 
mixing is very small, i.e. of order $wv/\mu u$.  Details of the phenomenology 
of this model will be discussed in a forthcoming paper.
\end{thebibliography}

\end{document}